A. Shipunov[1,2,3], A.K.H. Raghavendra[1], and G. Newcombe[1,2]

**Fungal endophytes in spotted knapweed influence its competitive interactions.**

[1]*Department of Forest Resources, University of Idaho, Moscow, ID 83844-1133, USA.*

[2]*Center for Research on Invasive Species and Small Populations, University of Idaho, Moscow, ID 83844-1133, USA.*

[3]*Department of Biology, Minot State University, Minot, ND 58707, USA.*

Author for correspondence: Alexey Shipunov. Tel.: (701) 858 3116; fax: (701) 858 3163. Email: dactylorhiza@gmail.com

**Abstract**     Fungal symbionts are often overlooked in studies of plant invasion. Nevertheless, their role could be essential to the competitive success of the invader. We studied fungal endophytes in the widespread invasive *Centaurea stoebe* (common knapweed). A preliminary experiment showed that endophytes in roots of *C. stoebe* significantly reduced the biomass of evolutionarily naïve neighbours (*Festuca idahoensis*), compared to endophyte-free *C. stoebe*. In the main experiment non-clavicipitaceous endophytes belonging to six phylotypes, were employed as root inoculants. Each of these endophytes again reduced the growth of naïve neighbours (*F. idahoensis*); and remarkably, each also increased the growth of adapted neighbours (*F. ovina*) that were tested for the first time. Four of the six endophytes caused *C. stoebe* to gain a competitive advantage over its naïve neighbour that was significantly greater than the endophyte-free *C. stoebe* over that same neighbour. However, endophyte-free *C. stoebe* had no greater competitive advantage over *F. idahoensis* than it had over *F. ovina.* Therefore, plant-plant interactions were dramatically affected by the presence of endophytes in a way that would favor invasion.

**Key words**     *Alternaria*; *Centaurea* invasion; community ecology; competition; fungal endophytes.

## Introduction

In plant invasions, a primary challenge is to understand the superior competitive ability of a successful exotic plant. Typically a successful invader is both less competitive and less abundant in its native range; this range-dependent puzzle of invasiveness is central to invasion biology. Although the contribution to plant invasions of release from fungal pathogens is well known (Mitchell & Power, 1991), studies of the contributions of endophytes have been initiated only recently (Addy *et al*., 2005; Faeth *et al.*, 2004; Omacini *et al.*, 2006; Rudgers *et al.*, 2005; Rudgers & Orr, 2009).

Recently, Rodriguez and co-authors used symbiotic criteria to group fungal endophytes of plants in four classes (Rodriguez *et al*., 2009). Class 1 endophytes belonging to the Clavicipitaceae are well known to ecologists as grass symbionts (Clay, 1988), and the pioneering investigation showed that *Neotyphodium caenophialum* promotes plant invasions (Rudgers *et al*., 2005). However, the other three classes of endophytes are uninvestigated with respect to their roles in plant invasions. We have found considerable diversity among non-clavicipitaceous endophytes in *Centaurea stoebe*, the European plant invader in North America that is commonly known as "spotted knapweed" (Shipunov *et al*., 2008). All 92 sequence-based, fungal phylotypes were obtained from cultures of seed isolates of *C. stoebe*. Since endophytes in classes 3 and 4 are not transmitted vertically through seed, whereas Class 2 endophytes are (Rodriguez et al., 2009), endophytes from *C. stoebe* seed are presumed to belong to Class 2. These endophytes can colonize and affect biomass of both root and shoot systems of plants, but their effects on plant competitiveness and invasiveness are unknown. To date, we have determined the effects of only a few of the 92 endophytes on the growth of *C. stoebe* itself (Newcombe *et al*., 2009). An important question is whether endophytes improve the competitiveness of their hosts versus plants that they encounter in their invaded range.

Our purpose here was to determine whether the most common endophytes of *C. stoebe* influence its competitive interactions with two species of *Festuca*, grasses that co-occur with spotted knapweed in both native (*F. ovina*) and invaded (*F. idahoensis*) ranges. In order to determine the existence and magnitude of these putative interactions, we designed a set of experiments (preliminary and main) that involved inoculations of seedling roots of *C. stoebe* with endophytes followed by competition with either of *Festuca idahoensis* or *F. ovina*.

## Methods

### A. Selecting the most abundant phylotypes

Seedheads of *C. stoebe* were sampled in its invaded range (mostly Northwestern U.S.) and its native range (Middle and Eastern Europe, European Russia, North Caucasus and the Urals). In all, 102 sites were sampled (53 from the invaded range and 49 from the native range). In each site or population of *C. stoebe*, five plants were sampled, and from each plant, 20 seeds (i.e., achenes), for a total of 100 seeds per site and 10,200 seeds in all. Endophytes were isolated onto potato dextrose agar, PDA, from seeds following 'Method II' surface-sterilization (Schulz *et al*., 1993). Each isolate received its own 'Cultivation Identification Number' (CID – Table 1), and was assigned on the basis of morphology and ITS and Alt a 1 sequences to a phylotype of a fungal genus. Methods for extraction, amplification and sequencing of the nuclear 5.8S rRNA gene and the two flanking, ITS regions were as previously published (Ganley *et al*., 2004). As a proxy for recognizing fungal species, ITS sequences may be conservative because biological species may share the same sequence (Lieckfeldt & Seifert, 2000). Because undescribed species may be common among endophytic isolates (Froehlich & Hyde, 2004; Ganley *et al*., 2004; Hartnett *et al*., 1993; Shipunov et al., 2008), a sequence-based approach is increasingly employed in endophyte studies. For those endophytes of *C. stoebe* that could be assigned on the basis of ITS sequences to *Alternaria* and related genera, the Alt a 1 gene was also sequenced to provide additional discrimination of phylotypes (Hong *et al*., 2005). It is important to bear in mind that a single phylotype does not represent a clone; individuals belonging to the same phylotype here may differ genetically at loci that were not sequenced, and even more significantly they may differ biologically. In other words, variation within a phylotype is akin to intraspecific variation, as would be expected for a species proxy. To determine the most abundant phylotypes for experiments, relative abundances of endophytes were calculated on a phylotype basis, and then representative isolates were selected for the inoculations of the main experiment, described below. Sequence data were deposited in GenBank (http://www.ncbi.nlm.nih.gov/).

B. Competition experiments

1. Preliminary experiment

Endophyte status was determined by germinating field-collected, surface-sterilized seeds of *C. stoebe* on 1.5% water agar in Petri dishes. E+ (endophyte infected) seedlings were ones from which endophytic fungi that had been in the seeds grew out into the agar; the roots of these seedlings were examined under a dissecting microscope to directly observe tissue darkening associated with infection. E- (endophyte-free) seedlings did not yield endophytes. These seven-day-old seedlings were then transplanted first to trays and then to pots two weeks later. Five, two-week-old seedlings of *F. idahoensis* were planted around each seedling of *C. stoebe*. In total, we prepared forty standard 3.78 dm$^3$ pots (20 per treatment). In this experiment, endophytes represented a random sampling of endophyte diversity in *C. stoebe* (Shipunov et al., 2008), as they had not yet been assigned to phylotypes.

2. Main experiment

For the main experiment, we employed 10-day-old cultures of representative isolates of the most abundant phylotypes (see below) to inoculate roots of seedlings germinated from seeds of *C. stoebe* plants grown in greenhouse. We had previously observed that individual plants of *C. stoebe* always produced endophyte-free seeds in greenhouse conditions. The experiment was conducted with representatives of the three most common phylotypes from each of the native and invaded ranges of *C. stoebe*: 1) isolates or CIDs of phylotypes 'alt002b', 'alt002c' and 'alt002f' from the native range; 2) isolates of 'alt002b', 'cla063', and 'epi066' from the invaded range (Table 1). Each of the six isolates was inoculated into roots of seven-day-old seedlings of *C. stoebe* by placing seedling roots in contact with a live culture of a particular endophyte for 12 hours. Root tissue darkening associated with infection was again checked under a dissecting microscope. Roots of control seedlings were placed in contact with uninoculated culture medium (i.e., PDA) for the same duration. After two weeks in trays, seedlings of *C. stoebe* were planted in pots with two-week-old neighbours that were either seedlings of evolutionarily naïve *F. idahoensis*, or adapted *Festuca ovina* from the exotic and native ranges of *C. stoebe*, respectively. This experiment comprised 192 pots, given 12 replicates of each combination of treatment (12 by 6 by 2, or 144 pots) and neighbor including E- control pots (12 by 2, 24 pots); plus 12 replicates of each neighbor without *C. stoebe* (24 pots).

In both experiments, pots were filled with sterilized 'Sunshine' mix (Sun Gro Horticulture Inc., Bellevue, WA, USA). Seeds of *F. idahoensis* were obtained from the Wind River Seed Co., Manderson WY; seeds of *F. ovina* were obtained from Grasslands West, Clarkston, WA. Greenhouse conditions included a 16h day, with temperatures between 24 and 27 °C. Each experiment was run for 18 weeks, at which point *C. stoebe* plants had flowered. Aboveground biomass was harvested, oven-dried to constant weight, and then weighed. If endophytes could affect competition, then the 'competitive advantage' of knapweed over fescue, was expected to be enhanced by endophytes and therefore biomass of endophyte-infected knapweed could prevalent over the biomass of fescue more than biomass of endophyte-free knapweed. Statistical analyses were performed both with R and with Systat version 12. The K-S Test (Lilliefors) was used to test data distributions and Levene's Test was used to test for homogeneity of variances.

**3. Re-isolation experiment**

To determine whether inoculation resulted in infection, we attempted to re-isolate inoculants of two phylotypes (CID 63 and CID 120) three weeks post inoculations. E- seedlings were treated as in the main experiment (see above), and then left to grow in a sterile environment for 21 days. Then seedlings were surface-sterilized with 50% ethanol (5 min) and distilled water and placed on the PDA medium.

**C. Presence of endophytes in roots of *Centaurea stoebe* in the field**

Field-collected roots of *C. stoebe* were sampled for endophytes. Because initial sequence data revealed multiple fungal species present in root tissues of plants in the field near Potlatch, Idaho, leading to mixed populations of ITS amplicons, PCR products were cloned, and individual sequences obtained from cloned PCR amplification products. One to three microliters of mixed, unpurified, undiluted PCR product were ligated overnight at room temperature into pGEM-T Easy TA cloning vector (Promega) in 10-microliter ligation reactions, following the manufacturer's protocol. One microliter of the ligation mixture was used to transform competent JM 109 *E. coli* cells, which were plated in multiple concentrations on LB/ampicillin plates (100 micrograms/mL) containing X-gal and IPTG. Presumptive recombinant colonies containing the cloned PCR product were screened by PCR for presence of appropriate insert; for each candidate colony, a 30-microliter PCR reaction was prepared containing ITS 1 and ITS 4 primers, PCR conditions and concentrations as described elsewhere (Ganley *et al*., 2004). Sterile micropipette tips were touched briefly to the surface of the candidate colony, and then rinsed in the PCR reaction by pipetting up and down two to three times. Reaction tubes were then placed into a thermal cycler without further treatment, and PCR carried out as usual. Five-microliter aliquots of completed PCR reactions were run on 1% agarose gels to check for amplification. Those containing insert of appropriate size were directly sequenced.

**D. Endophytes in *Festuca* neighbors**

To be sure that endophyte effects in the experiments were not due to *Festuca* endophytes, 300 seeds of *F. idahoensis* and 100 seeds of *F. ovina* were checked for *Neotyphodium* and other endophytes following surface-sterilization, and isolation as described above.

## Results

**Competition experiments**

In the preliminary experiment, the dry biomass of *F. idahoensis* in E- and E+ pots averaged 3.08 g and 2.20 g, respectively, on an individual plant basis. Endophytes in *C. stoebe* were thus responsible for significantly reducing the biomass of neighbouring *F. idahoensis* ($p \ll 0.01$, $F = 19.67$, $df = 1$). The biomass of inoculated *C. stoebe* itself was significantly higher than endophyte-free *C. stoebe* ($p = 0.009$, $F = 7.21$, $df = 1$) as E+ and E- *C. stoebe* averaged 13.40 g and 9.75 g, respectively. In sum, in the preliminary experiment, endophytes in *C. stoebe* were exerting negative effects on *F. idahoensis*. However, since the preliminary experiment was conducted with uncharacterized endophytes, we wondered whether observed effects were representative of the most common endophytes that we had isolated from *C. stoebe*.

The main experiment was conducted with representative isolates of the most common endophytic phylotypes found in seeds of *C. stoebe* (Table 1), after relative abundances had been determined. As in the preliminary experiment, the biomass of *F. idahoensis* was reduced by endophytes in *C. stoebe* (Fig. 1). However, this experiment also contrasted evolutionarily naïve and adapted neighbours, *F. idahoensis* and *F. ovina*, from the invaded and native ranges of *C. stoebe*, respectively. These neighbours were both affected by endophytes in *C. stoebe* but in opposite ways (Fig. 1). Whereas endophytes of *C. stoebe* generally reduced biomass of the naïve neighbour, *F. idahoensis*, they increased biomass of the adapted neighbour, *F. ovina*. Three of six endophytes significantly reduced the biomass of neighbouring *F. idahoensis* when compared to the effect of E- *C. stoebe* on *F. idahoensis*: CIDs 120, 63, and 73 (Bonferroni-adjusted, pairwise comparison p values = 0.003, 0.032, and 0.013, respectively). The first CID, 120, was from the Eurasian range of *C. stoebe*, but 63 and 73 were both isolated in North America. The effect of the Eurasian CID432 on neighbouring *F. idahoensis* was marginally significant as well (p = 0.062). CIDs 2, Eurasian, and 66, North American, reduced the biomass of *F. idahoensis* also (Fig. 1), but not significantly.

In striking contrast, four of six endophytes significantly increased the biomass of neighbouring *F. ovina* when compared to the effect of E- *C. stoebe* on *F. ovina*: CIDs 2, 432, 63, and 66 (Bonferroni-adjusted, pairwise comparison p values = 0.009, 0.002, 0.05, and 0.000, respectively). The first two of these were isolated in the Eurasian range of *C. stoebe*, and the last two were both isolated in North America. CIDs 120, Eurasian, and 73, North American, increased the biomass of *F. ovina* also (Fig. 1), but not significantly. Thus, the only *C. stoebe* endophyte to both significantly reduce the biomass of *F. idahoensis* and significantly increase that of *F. ovina* was CID 63, a *Cladosporium* isolate from North America.

Four of six endophytes caused *C. stoebe* to gain a competitive advantage over *F. idahoensis*, that was significantly greater than the competitive advantage of endophyte-free *C. stoebe* over *F. idahoensis*. These four endophytes were CIDs 2 (p = 0.01), 432 (p = 0.004), 63 (p = 0.03), and 73 (p = 0.001). Interestingly, CID 2 significantly increased competitive advantage of *C. stoebe* even though it had not significantly reduced biomass of *F. idahoensis*. Conversely, CID 120 did not significantly increase competitive advantage of *C. stoebe* over *F. idahoensis* even though it had significantly reduced biomass of *F. idahoensis*. The endophyte-free controls showed the lowest mean competitive advantage over *F. idahoensis* at 4.9 g (Table 2). Thus, CID73, the isolate of the 'alt002b' phylotype from North America, increased by over four times the competitive advantage of *C. stoebe* over *F. idahoensis* when compared to the endophyte-free control (21.3 g versus 4.9 g, respectively – Table 2).

Since four of six endophytes significantly increased the biomass of neighbouring *F. ovina* when compared to the effect of E- controls, one would expect an endophyte-mediated reduction in competitive advantage of *C. stoebe* over *F. ovina*. However, only CID 63 significantly reduced competitive advantage over *F. ovina* ($p = 0.03$) to -3.8 g per pot (Table 2). Even though CIDs 432 and 66 had significantly increased the biomass of *F. ovina*, each increased, though insignificantly, the competitive advantage of *C. stoebe* over *F. ovina*, when compared to the E- control.

Finally, *C. stoebe* gained a greater competitive advantage over its naïve neighbour, *F. idahoensis*, than that which it gained over its adapted neighbour, *F. ovina*, only when inoculated with endophytes: Biomass $_{\text{Endophyte-infected } C.\ stoebe}$ - Biomass $_{F.\ idahoensis}$ > Biomass $_{\text{Endophyte-infected } C.\ stoebe}$ - Biomass $_{F.\ ovina}$. Endophyte-free *C. stoebe* actually showed comparable competitive advantages over *F. idahoensis* and *F. ovina* (4.9 g versus 7.8 g, respectively – Table 2). In contrast, five of the six endophytes significantly increased the competitive advantage of *C. stoebe* over the naïve neighbour when compared to the advantage over the adapted neighbour (Table 2). The one exception was CID66, an *Epicoccum* isolate that did not cause a significant increase in competitive advantage over *F. idahoensis*, when compared with the endophyte-free controls (7.1 g versus 4.9 g, respectively – Table 2).

Biomasses of *C. stoebe* and *Festuca* species were inversely correlated for both *F. ovina* (Pearson $r = -0.40$; $p < 0.001$) and for *F. idahoensis* (Pearson $r = -0.41$, $p < 0.001$), as one might expect for moderate competition within pots. However, it was only when *C. stoebe* was growing with *F. idahoensis*, that biomass of *C. stoebe* was highly correlated with competitive advantage of the former over the latter (Pearson $r = 0.82$, $p < 0.001$). In contrast, there was no correlation between biomass of

*C. stoebe* and competitive advantage over *F. ovina* (Pearson r = 0.07, p = 0.51), largely because only CID 63 significantly affected competitive advantage over *F. ovina*, as discussed above.

The endophyte factor, with seven levels (i.e., six isolates plus the E- control), by itself explained 31% of the variation in competitive advantage over *F. idahoensis* (GLM; F = 5.76, p < 0.001). However, interaction between *C. stoebe* biomass and the endophyte factor actually explained slightly more variation, 36%, in competitive advantage over *F. idahoensis* (GLM; F = 7.36, p < 0.001) than endophytes alone. For both *F. idahoensis* and *F. ovina*, competitive advantage of *C. stoebe* was not as well explained by the interaction of endophytes with *Festuca* (i.e., biomass) as by the interaction of endophytes with their host, *C. stoebe* (biomass). *C. stoebe* biomass was itself significantly affected by endophyte treatments (GLM; F = 6.31, p < 0.001), as it had been in the preliminary experiment. Biomass of *F. idahoensis* grown by itself (i.e., five plants per pot), without *C. stoebe*, was significantly less than that of *F. ovina* grown by itself (p < 0.001, *t* = -4.17, df = 57).

**Re-isolation experiment**

Inoculants (i.e., CIDs 63 and 120) were commonly re-isolated indicating that infection had taken place. In several cases, we obtained isolates from plant tissues formed after inoculation, indicating that further colonization occurred after infection.

**Presence of endophytes in roots of *Centaurea stoebe* in the field**

Seed endophytes clearly had significant effects when inoculated into roots of *C. stoebe* plants in greenhouse experiments. But, did seed endophytes occur naturally in roots of *C. stoebe* in the field? Our sampling was not extensive but following cloning, all colonies with insert were sequenced, revealing four ascomycetous fungi: 1) a fungus with an ITS sequence identical to an "uncultured ascomycete clone", EU003079, in GenBank; 2) a fungus identical to *Protoventuria alpina*, EU035444 (Crous *et al*., 2007); 3) a fungus identical to an uncultured, soil fungus from the humic horizon,

EF434053 (Taylor *et al*., 2007); and 4) the 'cla063' phylotype that is the third most common seed endophyte of *C. stoebe* in its invaded range (Shipunov *et al*., 2008), and the endophyte that significantly reduced and increased biomasses of *F. idahoensis* and *F. ovina*, respectively, as reported here. With minimal sampling, 'cla063' was additionally found via cloning (i.e., the same approach used for detecting endophytes in roots) in leaves of *C. stoebe* in the field. This *Cladosporium* isolate, 'cla063', has thus been isolated from roots, leaves and seeds as one would expect for a Class 2 endophyte (Rodriguez et al., 2009).

**Endophytes in *Festuca* neighbours**

Surface-sterilized samples of the seed of *F. idahoensis* and *F. ovina* employed in the greenhouse experiments did yield some endophytes: four phylotypes from *F. idahoensis* and three from *F. ovina*. Isolation frequencies were thus low and approximately equal for the seed of *F. idahoensis* and *F. ovina* (i.e., 1.7% and 3%, respectively). *Neotyphodium* isolates, which are known to affect growth and interactions of *Festuca* (Van Hecke *et al*., 2005), were not obtained. There was no overlap (i.e., no endophytes in common) between the seven phylotypes from *Festuca* and the five phylotypes from *C. stoebe* of Table 1. The implications of these results in combination with results from the competition experiments suggest that influence of *Festuca* endophytes on experimental outcomes was minimal.

## Discussion

We found that competitive interactions between *C. stoebe* and its *Festuca* neighbours were affected by the presence of endophytes in *C. stoebe.* The identity of the neighbour mattered; effects on evolutionarily naïve *F. idahoensis* were negative, aiding *C. stoebe*, whereas effects on adapted *F. ovina* were positive. Our findings indicate that some of the endophytes of *C. stoebe* may increase its invasiveness, at least as gauged by competition with *F. idahoensis*. At a more general level, Class 2 endophytes should be considered an additional group of mutualistic agents that can promote plant invasions (Richardson *et al.*, 2000; Rudgers *et al.*, 2005).

The effects of endophytes were not tied to the range of *C. stoebe* (native or invaded) from which they were isolated; site of isolation does not by itself indicate the native range of an endophytic fungus (Shipunov *et al.*, 2008; Newcombe & Dugan, 2010). But just as the identity of the neighbouring *Festuca* species influenced competitive outcomes with *C. stoebe*, the identity of endophyte inoculants also mattered. For example, the *Epicoccum* isolate of the 'epi066' phylotype did not increase the competitive advantage of *C. stoebe* over the naïve competitor as compared to the adapted competitor (Table 2). Whereas the phylotype for which the evidence of Class 2 endophyte status was strongest (i.e., the *Cladosporium* isolate of the 'cla063' phylotype) did.

In our experiments, the roots of seedlings of *C. stoebe* were inoculated to mimic what appears likely to be a natural infection process following germination of endophyte-infected seed, and the re-isolation experiment showed that inoculation can result in infection.. Roots are more likely to be colonized systemically by endophytes than shoots (Boyle *et al*., 2001), but we do not yet know whether the effects reported here even depend on persistent root infection. Root turnover can provide a significant substrate for microbes in soil (Leigh *et al.*, 2002), and it is conceivable that endophytes alternate between *in planta* and soil phases. Endophytes might retard growth of naïve neighbours (Rudgers *et al*., 2005). Underground chemical compounds can be produced by invasive plants, as has been postulated for *C. stoebe* itself (Bais *et al*., 2003; Blair *et al*., 2005; Blair *et al*., 2006; Callaway & Aschehoug, 2000; Callaway & Ridenour, 2004; Vivanco *et al*., 2004), although this hypothesis is still controversial (Lau *et al*., 2008). Nutrient parasitism can also be mediated by mycorrhizal fungi (Carey *et al*., 2004), but the ascomycetous root endophytes employed here are not known to set up networks essential to this possible mechanism (Addy *et al*, 2005; Jumpponen, 2001). Barrier experiments coupled with observations of cleared and stained roots of both *C. stoebe* and its neighbors are needed.

Neighbour identity has been shown to affect plant interactions mediated by soil fungi (Callaway *et al*., 2003). Similarly, root inoculations with fungi have shifted coexistence ratios of *Populus* and invasive *Tamarix* in pot experiments (Beauchamp *et al*., 2005), and the roots appeared to be colonized mostly by dark septate endophytes that are likely ascomycetous as here. But, in the latter experiments also, mechanism remained unknown. Fungi can produce phytohormones (Tudzynski, 1997); in particular, *Alternaria* species can produce plant growth regulators (Kimura *et al*., 1992), and four of the six endophyte isolates employed here belonged to this genus. Mycorrhization can increase rates of net photosynthesis (Allen *et al*., 1981; Dosskey *et al*., 1990), but the endophytes employed in our main experiment are not known to do so. Alternatively, various rhizosphere microbes are also known to both up-regulate and down-regulate auxin activity in different plants (Ditengou & Lapeyrie, 2000), by acting on auxin-responsive genes such as *Pp-C61* (Reddy *et al*., 2003). Whatever their underlying mechanisms may be, the effects reported here suggest at the very least that endophytes may play important roles in plant community ecology, and their roles in plant invasions merit further study.

## Acknowledgements

Funding was provided by the Center for Research on Invasive Species and Small Populations of the University of Idaho. Mark Schwarzländer, Tim Prather, Linda Wilson, Melissa Baynes, Chandalin Bennett, Patrick Häfliger, and Heinz Müller-Schärer provided additional seedhead collections of *C. stoebe*. Tim Prather and Ray Callaway provided useful advice which helped us to improve the manuscript.

**Table 1.** The most common endophytic phylotypes of *Centaurea stoebe* in Eurasia and North America, on the basis of morphology and ITS and Alt a 1 sequences. Three, asterisked CIDs or isolates from each range that are representative of abundant phylotypes were used in experiments.

| Genus | Order | CID | Phylotype | GenBank accession [ITS sequence] | GenBank accession [Alt a 1 sequence] | Relative abundance in the native range, Eurasia | Relative abundance in the invaded range, North America |
|---|---|---|---|---|---|---|---|
| *Alternaria* | Pleosporales | 2 | alt002b | EF589849 | EF589830 | 43.54% * | - |
| *Alternaria* | Pleosporales | 73 | alt002b | EF589849 | EF589830 | - | 10.39% * |
| *Alternaria* | Pleosporales | 120 | alt002f | EF589849 | EF589833 | 6.08% * | 2.03% |
| *Alternaria* | Pleosporales | 432 | alt002c | EF589849 | EF589840 | 11.7% * | 0.1% |
| *Cladosporium* | Capnodiales | 63 | cla063 | EF589865 | - | 0.08% | 11.24% * |
| *Epicoccum* | Pleosporales | 66 | epi066 | EF589869 | - | 1.06% | 11.56% * |

CID: Cultivation Identification Number.

**Table 2.** Summary of effects of C*entaurea stoebe* endophytes: mean competitive advantage of *C. stoebe* over *Festuca idahoensis* versus advantage of *C. stoebe* over *F. ovina*.

| Endophyte | Neighboring *Festuca* species | Mean competitive advantage [*C. stoebe* biomass – *F. idahoensis* biomass], g (SE) | N | Neighboring *Festuca* species | Mean competitive advantage [*C. stoebe* biomass – *F. ovina* biomass], g (SE) | N | Pairwise comparison of means (Bonferroni-adjusted P) |
|---|---|---|---|---|---|---|---|
| CID120 | *F. idahoensis* | 15.6 (2.0) | 12 | *F. ovina* | 5.9 (2.0) | 12 | 0.008 |
| CID2 | *F. idahoensis* | 18.1 (3.6) | 12 | *F. ovina* | 3.9 (2.2) | 11 | <0.001 |
| CID432 | *F. idahoensis* | 19.3 (2.4) | 12 | *F. ovina* | 12.4 (2.7) | 12 | 0.06 |
| CID63 | *F. idahoensis* | 17.1 (3.0) | 12 | *F. ovina* | -3.8 (2.4) | 12 | <0.001 |
| CID66 | *F. idahoensis* | 7.1 (2.4) | 12 | *F. ovina* | 9.9 (2.8) | 12 | 0.445 |
| CID73 | *F. idahoensis* | 21.3 (2.0) | 12 | *F. ovina* | 5.6 (3.1) | 12 | <0.001 |
| Endophyte-free control. | *F. idahoensis* | 4.9 (2.4) | 12 | *F. ovina* | 7.8 (2.0) | 12 | 0.421 |

Figures:

**Figure 1.** Biomass of evolutionarily naïve *Festuca idahoensis* and adapted *F. ovina* affected by endophyte treatments of *C. stoebe* growing in the same pots. Treatmemnts reduced and increased biomass of *F. idahoensis* and *F. ovina*, respectively, when compared to their endophyte-free, or E-, controls. Endophyte isolates (CIDs 2, 63, 66, 73, 120, 432) represent the most common phylotypes in *C. stoebe*. Bars are means ± standard errors.

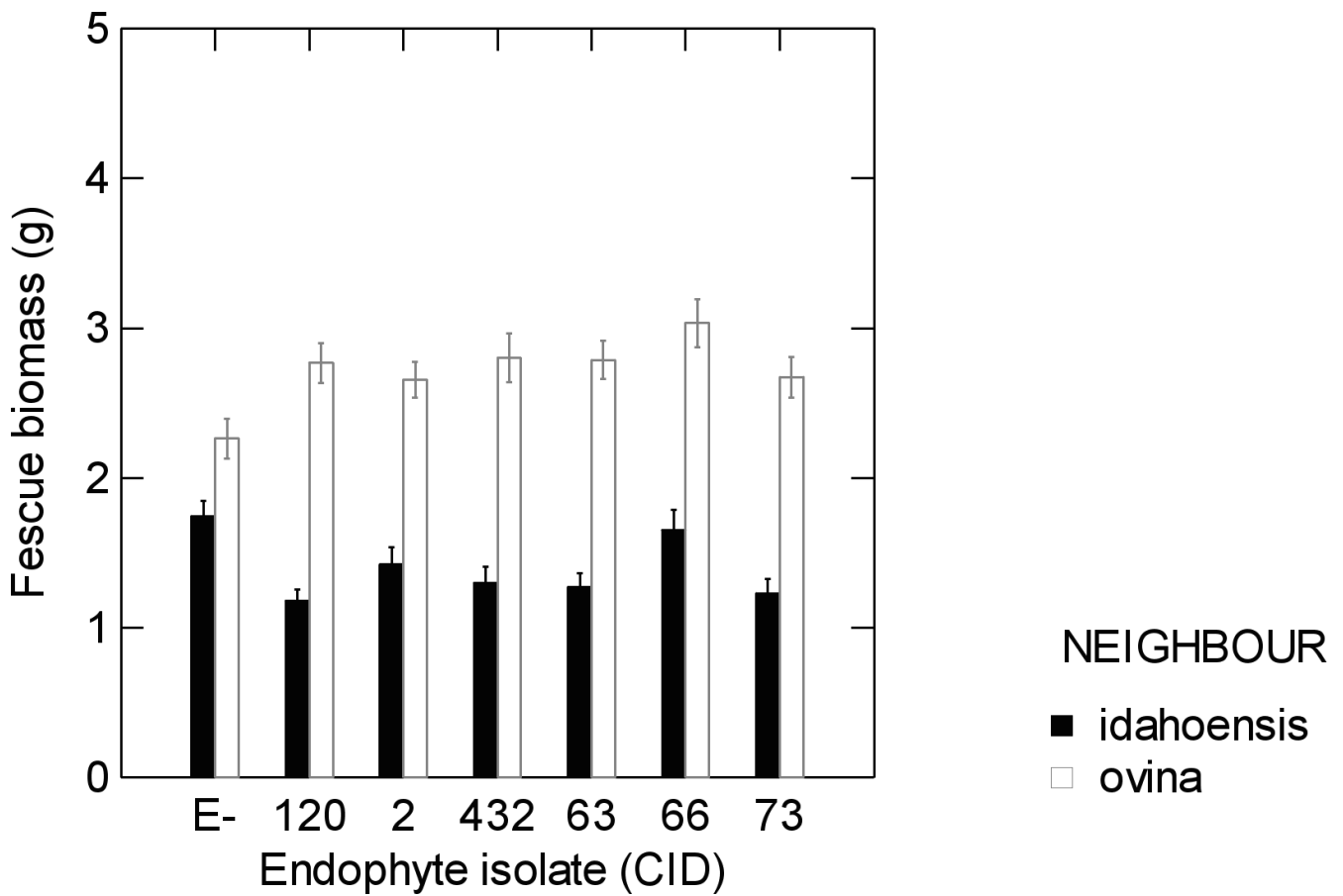